\documentstyle[12pt]{article}

\begin{document}
{\sf \begin{center} \noindent { \Large\bf  Stable curved solitonic surfaces in nonholonomic frame}\\[3mm]

by \\[0.3cm]{\sl L.C. Garcia de Andrade}\\

\vspace{0.5cm} Departamento de F\'{\i}sica
Te\'orica -- IF -- Universidade do Estado do Rio de Janeiro-UERJ\\[-3mm]
Rua S\~ao Francisco Xavier, 524\\[-3mm]Cep 20550-003, Maracan\~a, Rio de Janeiro, RJ, Brasil\\[-3mm]
Electronic mail address: garcia@dft.if.uerj.br\\[-3mm]
\vspace{2cm} {\bf Abstract}
\end{center}
\paragraph*{}
Assuming the stability of soliton surfaces of vanishing Ricci
sectional curvature of soliton metric in the nonholonomic frame, we
find a solution for the metric in the approximation of weak constant
torsion curves with constant Frenet curvature. The computation of
the Riemann tensor of the soliton metric shows that it does not
vanish and therefore the solution is nontrivial. Heisenberg
solitonic equation is also used to constrain the the soliton Riemann
metric. The new feature here is that the coordinate curves on the
soliton-like surface are composed of hydrodynamical
filaments.\vspace{0.5cm} \noindent {\bf PACS numbers:}
\hfill\parbox[t]{13.5cm}{02.40.Hw-Riemannian geometries}

\newpage
 \section{Introduction}
 Recently T. Kambe \cite{1} investigated the geometry and stability of solitons given by KdV equations of water
 waves as diffeomorphic flow, by computing the Riemann and sectional curvature , as well as the Killing vectors of the
 solitons. The stability of geodesic flows have been recently investigated by Kambe \cite{1} by making use of the
 technique of Ricci sectional curvature \cite{2}, where the negative
 sectional curvature indicates instability of the flow, while positivity or null indicates stability. In the case of
 instability the geodesics deviate from the perturbation of the fluid. In this letter, the
 sectional Riemann curvature of the geodesic flow for a Riemannian soliton surface \cite{3}, in the of nonholonomic
 frame, where the curves on the soliton surface possesses
 Frenet curvature and torsion, and where the legs of the nonholonomic frame, depend upon other coordinate directions
 orthogonal to the coordinate along one of the curves. In the some approximations and assuming the stability of the soliton
 surface metric along with Heisenberg solitonic constraint, we find the metric for this soliton surface for incompressible filamentary
 flows. A distinct aspect between Kambe´s and ours approach is that the sectional curvature does not depend on the flow
 speed but just on the geometrical quantities of the flow. This is similar to an idea of Thifeault \cite{4} which describes the
 Riemannian geometry of a curved substrate. The paper is organized as follows: Section II presents a
 brief review of Riemannian geometry in the coordinate free language.
 Section III presents the sectional curvature for the solitonic like surface. Section IV presents the conclusions.
 \section{Ricci and sectional Riemann curvatures}
 In this section we make a brief review of the differential geometry of surfaces in coordinate-free language.
 The Riemann curvature is defined by
 \begin{equation}
 R(X,Y)Z:={\nabla}_{X}{\nabla}_{Y}Z-{\nabla}_{Y}{\nabla}_{X}Z-{\nabla}_{[X,Y]}Z\label{1}
 \end{equation}
where $X {\epsilon} T\cal{M}$ is the vector representation which is
defined on the tangent space $T\cal{M}$ to the manifold $\cal{M}$.
Here ${\nabla}_{X}Y$ represents the covariant derivative given by
\begin{equation}
{\nabla}_{X}{Y}= (X.{\nabla})Y\label{2}
 \end{equation}
which for the physicists is intuitive, since we are saying that we
are performing derivative along the X direction. The expression
$[X,Y]$ represents the commutator, which on a vector basis frame
${\vec{e}}_{l}$ in this tangent sub-manifold defined by
\begin{equation}
X= X_{k}{\vec{e}}_{k}\label{3}
\end{equation}
or in the dual basis ${{\partial}_{k}}$
\begin{equation}
X= X^{k}{\partial}_{k}\label{4}
\end{equation}
can be expressed as
\begin{equation}
[X,Y]= (X,Y)^{k}{\partial}_{k}\label{5}
\end{equation}
In this same coordinate basis now we are able to write the curvature
expression (\ref{1}) as
\begin{equation}
R(X,Y)Z:=[{R^{l}}_{jkp}Z^{j}X^{k}Y^{p}]{\partial}_{l}\label{6}
\end{equation}
where the Einstein summation convention of tensor calculus is used.
The expression $R(X,Y)Y$ which we shall compute bellow is called
Ricci curvature. The sectional curvature which is very useful in
future computations is defined by
\begin{equation}
K(X,Y):=\frac{<R(X,Y)Y,X>}{S(X,Y)}\label{7}
\end{equation}
where $S(X,Y)$ is defined by
\begin{equation}
{S(X,Y)}:= ||X||^{2}||Y||^{2}-<X,Y>^{2}\label{8}
\end{equation}
where the symbol $<,>$ implies internal product.
\newpage
\section{Ricci stable solitonic surfaces} In this section we shall consider the metric $g(X,Y)$ of line element of solitonic surface
defined as the first fundamental form of differential form\cite{3}
\begin{equation}
I_{\sum}=ds^{2}+g(s,b)db^{2} \label{9}
\end{equation}
This is a Riemannian line element
\begin{equation}
ds^{2}=g_{ij}dx^{i}dx^{j} \label{10}
\end{equation}
where $g_{11}=1$ and $g_{22}=g(s,b)$. The curves or filaments
generating this solitonic like surface is given by the two vector
tangent fields
\begin{equation}
X=u_{s}(s)\vec{t} \label{11}
\end{equation}
and
\begin{equation}
Y=u_{b}(s)\vec{t} \label{12}
\end{equation}
obeying the imcompressible flow equation
\begin{equation}
{\nabla}.\vec{u}={\partial}_{s}u_{s}+g^{-\frac{1}{2}}{\partial}_{b}u_{b}(s)\vec{t}=0
\label{13}
\end{equation}
Let us now assume stability of the filamentary flows on the soliton
like surface embbeded in the Euclidean manifold ${\cal{E}}^{3}$, and
compute the Ricci sectional curvature above step by step. We need
first to make use of the grad operator in the  Riemannian solitonic
metric, which is given by
\begin{equation}
\nabla=[{\partial}_{s},g^{-\frac{1}{2}}{\partial}_{b},{\partial}_{n}]
\label{14}
\end{equation}
The non-holonomic dynamical relations from vector analysis and
differential geometry of curves \cite{5} such composed the Frenet
frame $(\vec{t},\vec{n},\vec{b})$ equations
\begin{equation}
\vec{t}'=\kappa\vec{n} \label{15}
\end{equation}
\begin{equation}
\vec{n}'=-\kappa\vec{t}+ {\tau}\vec{b} \label{16}
\end{equation}
\begin{equation}
\vec{b}'=-{\tau}\vec{n} \label{17}
\end{equation}
and the other frame direction legs are given by
\begin{equation}
\frac{\partial}{{\partial}n}\vec{t}={\theta}_{ns}\vec{n}+[{\Omega}_{b}+{\tau}]\vec{b}
\label{18}
\end{equation}
\begin{equation}
\frac{\partial}{{\partial}n}\vec{n}=-{\theta}_{ns}\vec{t}-
(div\vec{b})\vec{b} \label{19}
\end{equation}
\begin{equation}
\frac{\partial}{{\partial}n}\vec{b}=
-[{\Omega}_{b}+{\tau}]\vec{t}-(div{\vec{b}})\vec{n}\label{20}
\end{equation}
\begin{equation}
\frac{\partial}{{\partial}b}\vec{t}={\theta}_{bs}\vec{b}-[{\Omega}_{n}+{\tau}]\vec{n}
\label{21}
\end{equation}
\begin{equation}
\frac{\partial}{{\partial}b}\vec{n}=[{\Omega}_{n}+{\tau}]\vec{t}-
\kappa+(div\vec{n})\vec{b} \label{22}
\end{equation}
\begin{equation}
\frac{\partial}{{\partial}b}\vec{b}=
-{\theta}_{bs}\vec{t}-[\kappa+(div{\vec{n}})]\vec{n}\label{23}
\end{equation}
Therefore to compute the Ricci tensor step by step we start by the
second term in the Ricci tensor is
\begin{equation}<{\nabla}_{X}{\nabla}_{Y}Y,X>=-{u_{s}}^{2}{u_{b}}^{2}{\kappa}_{0}g^{-\frac{1}{2}}[{\kappa}+div(\vec{n})]
\label{24}
\end{equation}
where we have used the helical filaments hypothesis
${\kappa}_{0}=constant={\tau}_{0}$. The other terms in the Ricci
tensor are
\begin{equation} [X.Y]=-{u_{s}}{u_{b}}[1-g^{-\frac{1}{2}}]\vec{n}\label{25}
\end{equation}
which implies that
\begin{equation} {\nabla}_{[X,Y]}Y=-u_{s}u_{b}[1-g^{-\frac{1}{2}}]\vec{n}
\label{26}
\end{equation}
In these last equations we have used the Heisenberg constraint
equation \cite{5}
\begin{equation} {\partial}_{s}{\kappa}={\theta}_{bs}\kappa\label{27}
\end{equation}
which allows us to place ${\theta}_{bs}=0$ in future computations.

\newpage

The sectional curvature is thus
\begin{equation}
K(X,Y)= \frac{<R(X,Y)Y,X>}{S(X,Y)}=
-[1-g^{-\frac{1}{2}}]{\theta}_{ns}+{{\tau}_{0}}^{2}-g^{-\frac{1}{2}}[{{\tau}_{0}}^{2}+{\tau}_{0}div\vec{n}]\label{28}
\end{equation}
Since it is assumed that the solitonic surface is stable we simply
place $K(X,Y)=0$ in the last expression, which yields the following
solution for the solitonic metric
\begin{equation}
g(s,b)=[{1+\frac{div\vec{n}}{{\tau}_{0}}}]^{2}\label{29}
\end{equation}
when the filaments possess a very weak torsion we can approximate
this expression by
\begin{equation}
g(s,b)\approx{[\frac{div\vec{n}}{{\tau}_{0}}]^{2}}\label{30}
\end{equation}
Note that when both angular velocity and perturbation both keep the
same sign, the sectional curvature $K(X,Y)$ is negative and the flow
along the Ricci soliton is stable.
\section{Conclusions}
An important issue in plasma astrophysics as well as in fluid
mechanics and optics is to know when a fluid, charged or not, is
unstable or not. In this we invert the problem and assume stability
of solitonic-like surfaces and found with the appropriated
constraints the metric for this soliton surface. Incompressible
flows filaments are used in the formation of the curved surface.

\end{document}